\documentclass[11pt]{article}
\usepackage{graphicx}

\newsavebox{\PSLASH}
\sbox{\PSLASH}{$p$\hspace{-1.8mm}/}

\begin{document}
\title{Conformal Curves in Potts Model: Numerical Calculation }
\author{F. Gliozzi$^a$\footnote{e-mail: gliozzi@to.infn.it} and   M. A. Rajabpour$^{a}$\footnote{e-mail: Rajabpour@to.infn.it} \\ \\
 $^{a}$Dip. di Fisica Teorica and INFN, Sezione di Torino,\\
Universit{\`a} di Torino, Via P. Giuria 1, 10125 Torino, Italy
}
\maketitle
\begin{abstract}
We calculated numerically the fractal dimension of the boundaries of the Fortuin-Kasteleyn clusters of
the $q$-state Potts model for integer and non-integer values of $q$ on the square lattice.
 In addition we calculated with high accuracy the fractal dimension of the boundary points of the same clusters on the
square domain. Our calculation confirms that this curves can be described by SLE$_{\kappa}$.   

\vspace{5mm}
%\newline
\end{abstract}

\section{Introduction}
Studying the critical interfaces in two dimensions is in the
interest of both physicists and mathematicians because of at least
two reasons: firstly because of application in the physical systems
such as domain walls in statistical models \cite{bernard0,hastings},
iso-height lines in rough surfaces \cite{rajab}, zero-vorticity
lines of Navier-Stokes turbulence \cite{bernard} and nodal lines of random wave functions \cite{Nodal line}. Secondly these
interfaces when they posses conformal symmetry can be studied by the
exact methods of conformal field theory (CFT) \cite{BPZ} and rigorous
methods of Schramm-Loewner evolution (SLE) \cite{Schramm}.

Although  physicists introduced many interesting results by
considering the conformal invariance of the interfaces there is no
rigorous proof for most of them. It was just recently that
mathematicians were able to prove in some cases the conformal
invariance of the statistical models at the critical point by using
SLE techniques. Using SLE techniques Smirnov proved \cite{smirnov1}
the conformal invariance of percolation clusters which was conjectured long
time ago by Cardy \cite{Cardy} and checked numerically by Langlands
\textit{et al.} \cite{langland}\footnote{These authors checked numerically the invariance
 under M\oe bius transformation and not invariance under the whole 
conformal maps.}. The proof was based on site
percolation on the triangular lattice. There were lots of attempts
to find a similar proof for other lattices but up to now the
solution is out of reach \cite{beffara2}. In 2007 again Smirnov
proved the conformal invariance of Ising clusters in the
Fortuin-Kasteleyn (FK) representation on the square lattice
\cite{smirnov2}. Finally recently Chelkak and Smirnov proved the
conformal invariance of spin clusters of Ising model on the generic
iso-radial graph \cite{Smirnov3}. Apart from the above list of
rigorous proofs of well-known statistical models there are just three more proofs for conformal
invariance of stochastic curves, loop erased random walk \cite{LSW},
harmonic explorer \cite{SS0} and the contour lines of Gaussian free
field theory \cite{SS}.

Potts model is one of the building blocks of statistical mechanics
and it was studied from many different points of view. In two-dimensions at the self dual points it is an exactly solvable model and the critical 
properties were studied in detail, see \cite{Baxter}. 
It was also studied from CFT point of view
\cite{Difrancesco}. The critical domain walls in Potts model were
also studied in detail by using Coulomb gas techniques and lots of
conjectures came out from heuristic arguments \cite{SD,coniglio}. Some
numerical calculation were done to support the conjectures for the
Potts models with integer $q$, see \cite{Mandelbrot,WW,janke,PRS,ASZ}. They
mostly calculated the fractal dimension of domain walls, cluster
masses, red bonds, fjords and singly connected bonds, by simulating
them on the square lattice in the bulk of the rectangular domain. Although none
of them separately guarantees that those curves are SLE and so
conformal since there are some relations between the above
quantities and $\kappa$ of SLE for conformal curves \cite{janke}, one
can get convinced that they are related to SLE after checking at
least the validity of two of them. In other words if two of the
above quantities satisfy specific relation between each other then
the curves are conformal. The above argument still is not enough to
rule out for example the possibility of having SLE$(\kappa,\rho)$,
see \cite{LSW2}. To check that a process is SLE$(\kappa)$ or
SLE$(\kappa,\rho)$ one needs to check directly the properties of the drift of
SLE or to check some boundary fractal properties. It is difficult to
get some results with high accuracy by working directly with SLE
equation \cite{TK}. However, it is easier to calculate the boundary
fractal dimension of the curves in simulation. 

By the above
motivations in this article we will give some strong non-direct
simulation supports for the conformal invariance of the boundary of
FK clusters in the Potts model for integer and non-integer values of
$q$. To the best of our knowledge for the non-integer case there is
no simulation or rigorous argument for having conformal domain
walls. To rule out the possibility of having SLE$(\kappa,\rho)$ we
will calculate the boundary fractal dimension of the curves.

\section{Definition and Theory}
The definition of $q$-state Potts model on the square lattice in
the arbitrary domain is as follows: associate a spin variable
$s_{i}\in \{0,1,...,q-1\}$ at each site then the partition function
is
\begin{eqnarray}\label{Partition function}
Z=\sum_{s_{i}}  e^{J\sum_{<i,j>}\delta(s_{i},s_{j})}=\sum_ {s_{i}}
\prod_{<i,j>}(1+u\delta(s_{i},s_{j})),
\end{eqnarray}
where $u=e^{J}-1$. By expanding the product one can obtain the FK representation as follows
\begin{eqnarray}\label{FK}
Z=\sum_{G}u^{b}q^{c},
\end{eqnarray}
where $G$ is any subgraph of the original domain, consisting of all
the sites and some bonds placed arbitrarily on the lattice edges,
$b$ is the number of bonds in $G$, and $c$ is the number of clusters
of connected sites into which the bonds partition the lattice.
Although the original formulation of the model requires $q$ to be a
positive integer, the FK representation allows one to interpret $q$
as taking arbitrary real values. The model is known to have a
critical point at the self-dual value of $u_{c}=\sqrt{q}$ for $0\leq
q\leq4$. The partition function
 can be also expressed as that of a gas of fully packed loops on
the medial lattice. At the critical point FK clusters are equivalent
to counting each loop on the medial lattice with a fugacity $u_{c}$,
see \cite{Nienhuis}. It was conjectured in \cite{RS} that these
loops can be described by SLE with the following equation
\begin{eqnarray}\label{kappa}
\kappa=\frac{4\pi}{\cos^{-1}(-\frac{\sqrt{q}}{2})},
\end{eqnarray}
where $4\leq \kappa \leq 8$. Since the fractal dimension of SLE is
\begin{equation}
d_f=1+\frac{\kappa}{8}~, 
\label{df}
\end{equation}
(see \cite{beffara}), then one can find the
following formula for the fractal dimension of loops in the q-state
Potts model
\begin{eqnarray}\label{fractal potts}
\sqrt{q}=-2\cos(\frac{\pi}{2(d_f-1)}).
\end{eqnarray}
This is a well known formula in physics literature from the Coulomb
gas arguments \cite{coniglio} but it has not been checked numerically for
 non-integer values of $q$. In the next section we will check the
validity of the above equation for some values of $q$ in the
rectangular domain. The popular version of SLE is defined on the upper half
plane, a stochastic curve emerges from origin and goes to infinity
by keeping the right-left symmetry. Since SLE is a conformal curve, mapping 
the upper half plane to rectangle does not change the fractal
properties of the curve; for the non-conformal curves this argument
is not true. With the above argument one should be careful that
checking the equation (\ref{fractal potts}) on the rectangular
domain does not say anything about the conformality of the curve. As
we argued in the introduction checking at least one more quantity is
necessary for stronger numerical argument. One interesting quantity is the
fractal dimension of the boundary points of the fractal curve. It
was proved in \cite{boundaryfractal} that the fractal dimension of
the SLE points on the real line is
\begin{equation} 
d_b=2-\frac{8}{\kappa}~,
\label{db} 
\end{equation}
then one can find the following formula for the fractal dimension of
boundary points for the q-state Potts model loop
\begin{eqnarray}\label{boundaryfractal}
d_b=1-\frac{\cos^{-1}(-\frac{\sqrt{q}}{2})}{\pi}.
\end{eqnarray}
Checking the above equation in the rectangular domain is a stronger
numerical check for conformality of the stochastic curves. The above
equation is also important to support that these curves are related
to SLE$(\kappa)$ and not to SLE$(\kappa,\rho)$. The latter one is also a
conformal curve and has similar bulk properties but the fractal
dimension of the boundary points is different \cite{SJD}. The most
famous statistical model related to SLE$(\kappa,\rho)$ is the
contour lines of Gaussian free field theory \cite{SS}.

We will close this section by a short comment on the relation of the bulk and boundary fractal dimensions to the 
conformal field theory operators. From  CFT and Coulomb gas arguments it is well known that $\phi_{0,\frac{n}{2}}$ 
is related to the point with $n$ attached curves. Since the bulk fractal dimension is related to $n=2$ then the responsible 
bulk field is $\phi_{0,1}$ with weight $2h_{0,1}=1-\frac{\kappa}{8}$. Then the fractal dimension is $d=2-2h_{0,1}$. 
To produce a boundary point with $n$ attached curves one needs to plug the operator $\phi_{1,n+1}$ at the corresponding point. 
Then the fractal dimension of boundary points is $d_b=1-h_{1,3}$, where $h_{1,3}$ is the weight of the operator $\phi_{1,3}$. 

Combining equations (\ref{df}) and (\ref{db}) one obtains the relation
\begin{equation}
\left(d_f-1\right)\left(2-d_b\right)=1~,
\label{dfb}
\end{equation}
where any specific reference to the q-state Potts model has disappeared, therefore one could suspect that such a relation between the Hausdorff dimensions $d_f$
and $d_b$ could be true for a wider class of fractal curves.

In the next section we will give some extensive numerical simulations to check the validity of the equations (\ref{fractal potts}), (\ref{boundaryfractal})
and{(\ref{dfb}).

\section{Simulations}
\begin{figure} [htb]
\centering
\includegraphics[width=0.55\textwidth]{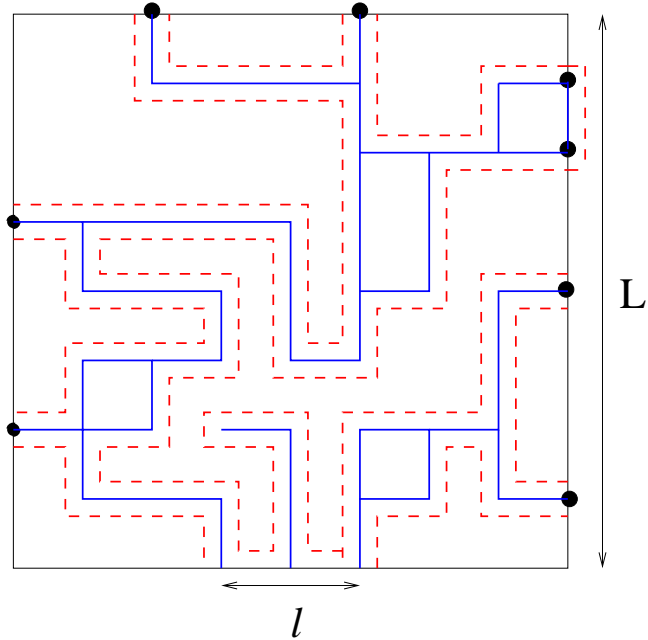}
%\vspace{5cm}  % amount of vertical space needed
\caption{Setting of our numerical simulations: in a square box of size 
$L\times L$ we measured the sum of the perimeters (dashed line)of all the FK 
clusters (thick lines) touching a fixed segment of the boundary of length 
$\ell=L/4$ placed symmetrically at the centre of one side. 
We also evaluated  the fractal dimension of the boundary by counting the number of the boundary sites (black dots) touched by these clusters.}
\label{Figure:1}
\end{figure}

We simulated three different systems in a square lattice: a bond percolation 
system at the threshold, corresponding to a $q=1$ Potts model with $\kappa=6$,  
a critical $q=2$ Potts associated to $\kappa=\frac{16}3$ and finally a 
critical $q=2\cos^2(\frac\pi5)=\frac{\sqrt{5}+3}2$ Potts model which 
according to (\ref{kappa}) corresponds to $\kappa=5$. 
 In all cases the system was enclosed in a  square box of 
size $L\times L$. We measured the sum of the perimeters of all the FK clusters 
touching a segment of length $\ell=L/4$ placed symmetrically on the centre of a side of the box (see Figure \ref{Figure:1}). 

In the case of percolation the boundary conditions were chosen free along the whole perimeter $B$  of the square box. 
 We adopted an epidemic type of algorithm: to begin the 
process, all the sites of the lattice are set to the ``unvisited'' state 
and bonds to the ``undetermined'' state. Then we pick one of the unvisited 
sites $i$ of the segment $\ell$ and we check the three bonds connected to 
$i$; the undetermined bonds are set ``occupied'' with probability 
$p=p_c=\frac12$ and ``empty'' otherwise. 
For bonds that are occupied, we check the adjacent site; if that site 
is unvisited we label it as visited and put it in a list for further 
checking. After finishing checking all bonds connected to a site, one considers
 the next site on the list, continuing this process until the list is empty. 
One repeats this process for each unvisited site of the segment $\ell$. 
In this way we generate only the clusters emanated by $\ell$. 
This algorithm allows to consider boxes of large size with little 
computational effort. In each sample generated this way we measured the 
sum of the perimeters of these  clusters as well as the number of sites of the 
boundary reached by these clusters.  We generated  two sets of data; in the first set we considered lattices of size $L\times L$ with $L=80,120,\dots,720$ with $1.2\,10^7$ samples for 
each $L$. The second set was composed of larger boxes 
$(L=1360,1600,1800,2000,2400,\dots,4000)$ with about $6\,10^6$ samples for each $L$. Figure \ref{Figure:2} shows the results for a plot of the total perimeter of the FK clusters versus $L$ in a logarithmic scale
\begin{figure}
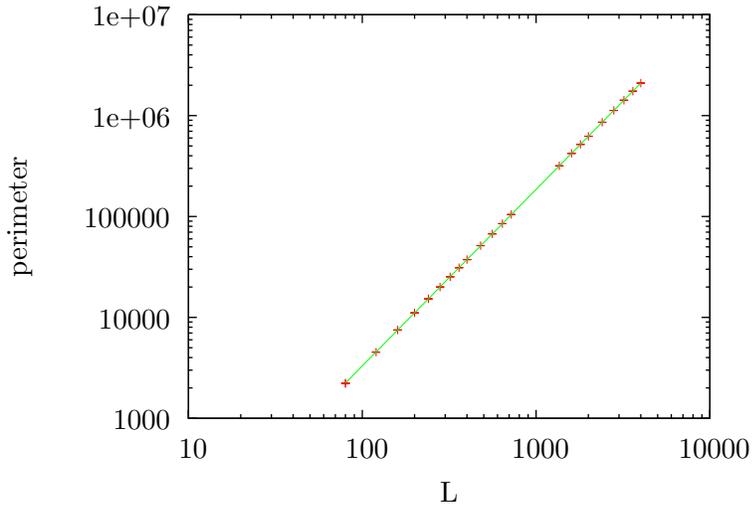

\centering
\input Slefig/dfperc
\caption{Plot of the total perimeters of the clusters emanated by the segment
$l$ as a function of the size $L$ of the percolation process. The data lie 
on a straight line in this logarithmic plot. The slope yields an estimate of the fractal dimension $d_f=1.74990\,\pm0.00002$ in perfect agreement with the 
expected exact value $d_f=1+\frac34$.}
\label{Figure:2}
\end{figure} 

In the cases of Potts models with $q>1$ we performed two different kinds of 
Monte Carlo simulations. For the estimate of the fractal dimension $d_f$ 
of total perimeter of the 
FK clusters emanated by the segment $\ell$ we chose fixed boundary conditions 
along this  segment, say $s_i=0~\forall\,i\in\ell$, and   
$s_i\not=0~\forall\,i\in\,B\setminus\ell$, while in estimating the Hausdorff dimension $d_b$ of the boundary we chose always fixed boundary conditions along 
$\ell$, but free boundary conditions along the rest $B\setminus\ell$ of the 
square box.
\begin{figure}
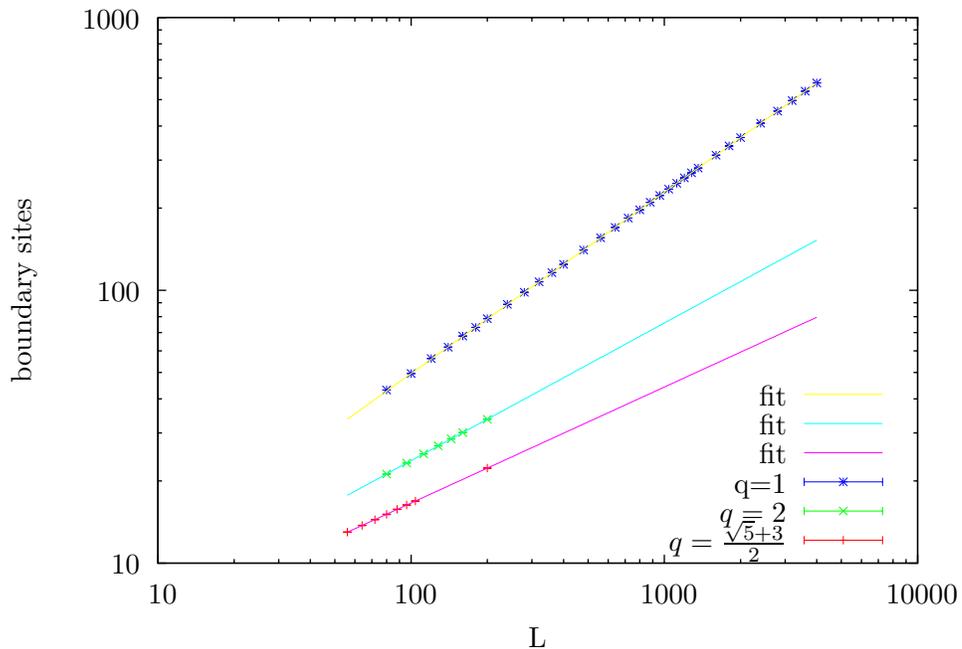

\centering
\input Slefig/boundary
\caption{Plot of the mean number of boundary sites attained by the FK clusters 
in the three Potts models studied. The three continuous lines are the fits of the data to $a\,L^{d_b}$. The size of the systems with $q>1$ is much smaller than in the case of percolation causing larger statistical errors in the determination of $d_b$, as Table 1 shows.}
\label{Figure:3}
\end{figure}
While these boundary conditions have an obvious implementation in the $q=2$ case
where we adopted a standard  Swendsen-Wang algorithm \cite{sw}, they need further specification in the case of non integer $q$, where the spin variables 
$s_i$ are ill-defined. Among the proposed algorithms for non integer $q$ 
\cite{sweeny,cm,gl}, the most efficient one for the $q>1$ case is that of 
Chayes and Machta, which  we implemented in the following form, starting from an arbitrary configuration of  clusters of connected sites

\begin{enumerate}
\item set all clusters to the ``undetermined'' state;
\item assign  to each undetermined cluster a white color with 
probability $\frac1q$ and a non-white color otherwise;
\item set each bond between white sites ``occupied'' with probability 
$\frac u{1+u}$ and ``empty'' otherwise. In this way a new configuration of clusters of connected sites is generated;
\item  return to step 1.
\end{enumerate}

\begin{table}[htb]
\caption{Estimated fractal dimension $d_f$ and $d_b$ in the critical Potts models with $q=1$, $q=2$ and $q=\frac{\sqrt{5}+3}2$ compared with the expected value
in $SLE$. }
  \begin{center}
\begin{tabular}{|c|c|c|c|c|}
\hline
$q$&$d_f$&expected $d_f$& $d_b$& expected $d_b$\\
\hline
1&$1.74990 \pm0.00002$&$\frac74$ &$0.6656\pm0.0002$&$\frac23$\\
2&$1.659\pm0.004$&$\frac53$&$0.505\pm0.003$&$\frac12$\\
$\frac{\sqrt{5}+3}2$&$1.625\pm0.003 $&$\frac{13}8$&$0.41\pm0.01$&$\frac25$\\
\hline
\end{tabular}
\end{center}
\end{table}

Note that  this algorithm uses only two kinds of sites - white and non-white -
but they are not treated equivalently. In our numerical estimates of $d_f$ we
chose white sites in the segment $\ell$ and non-white sites  in the rest of the boundary $B\setminus \ell$, while in the case of the determination of the fractal dimension $d_b$ the boundary no restriction was made on the sites of 
$B\setminus\ell$.

In order to extract our numerical estimates of the fractal dimension  we 
fitted our data to the power law $a\,L^d$ using $a$ and $d$ as fitting 
parameters. Since this is an asymptotic expression, valid when 
the size $L$ of the box is much larger than the lattice spacing, we fitted 
the data to the power law by progressively discarding the short distance 
points until the $\chi^2$ test gave a good value. The estimated values are reported in Table 1 and in Figure \ref{Figure:4}.
\begin{figure}
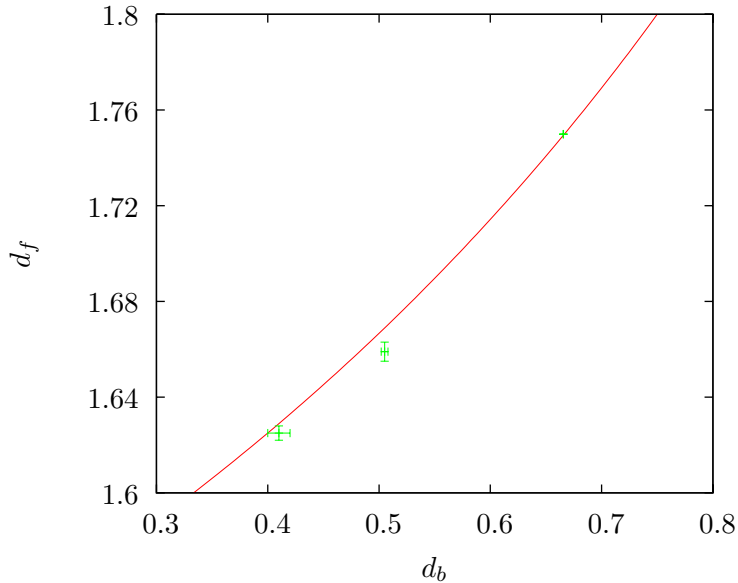

\centering
\input Slefig/dfdb
\caption{Plot of the estimated fractal dimensions in the plane $(d_b,d_f)$. 
The continuous curve represents Eq.(\ref{dfb}).}
\label{Figure:4}
\end{figure}
\section{Conclusions}
In this paper we calculated numerically the fractal dimension of the boundaries of the FK clusters of the Potts model, 
for the integer and non-integer values of $q$,
on the
square lattice. In addition we found precisely the fractal dimension of the boundary touching points of the FK clusters on the square lattice. Since
 the bulk fractal dimension and the boundary fractal dimension of the  domain walls of FK clusters of the Potts model have the same relations
that we expect from SLE we  believe that our method gives convincing numerical
 support for the conformal invariance of the boundaries of the FK clusters. The direct method to see the relation between
the critical curves and SLE
is calculating the drift of the Loewner equation and checking its equality, in the distribution sense, with the Brownian motion. The numerical methods in this direction is not
that much efficient and one can not get very precise numbers. For this reason checking at least two fractal properties of the curves can give
more efficient numerical support for the conformal invariance of the model.
\newline
\newline
\textit{Acknowledgment}: We are indebted to M. Caselle and S. Lottini for fruitful discussions.
\newline
\newline
\textit{Not added}: After completion of the present work, the paper \cite{ASZ} has appeared. It also calculates the fractal dimension
of the boundary of FK clusters with the comparable accuracy for integer $q$'s.

\end{document}